\documentclass[
aps,
prb,
reprint,
longbibliography,
floatfix,
superscriptaddress
]{revtex4-2}

\usepackage{graphicx}
\usepackage{bm}
\usepackage{amsmath}
\usepackage{amssymb}
\usepackage[
colorlinks=true,
citecolor=blue,
linkcolor=blue,
urlcolor=blue,
pdftitle={Engineering Competing Fractional Topological States in a BHZ Superlattice},
pdfauthor={Ke-Chen Liu and Chen Cheng}
]{hyperref}

\graphicspath{{./}}

\newcommand{\FirstAuthorName}{Ke-Chen Liu}
\newcommand{\CorrespondingAuthorName}{Chen Cheng}
\newcommand{\CorrespondingAuthorEmail}{chengchen@lzu.edu.cn}
\newcommand{\LZUTheoryAffiliation}{%
Lanzhou Center for Theoretical Physics, Key Laboratory of Quantum Theory
and Applications of MoE, Key Laboratory of Theoretical Physics of Gansu Province,
and Gansu Provincial Research Center for Basic Disciplines of Quantum Physics,
Lanzhou University, Lanzhou, Gansu 730000, China}

\begin{document}

\title{Engineering Competing Fractional Topological States in a BHZ Superlattice}

\author{\FirstAuthorName}
\affiliation{\LZUTheoryAffiliation}
\author{\CorrespondingAuthorName}
\email{\CorrespondingAuthorEmail}
\affiliation{\LZUTheoryAffiliation}

\begin{abstract}
Scalar superlattices offer a route from simple topological bands to
fractionalized states, while multicomponent systems raise the additional
question of which fractional order is energetically selected. We show that a
scalar superlattice applied to the Qi-Wu-Zhang (QWZ) Chern-insulator model
produces an isolated \(C=+1\) miniband with strongly improved Berry-curvature
and quantum-metric uniformity. Exact diagonalization of the projected
interaction identifies a \(1/3\) fractional Chern insulator in this miniband.
Restoring the time-reversed partner to form the Bernevig-Hughes-Zhang (BHZ)
superlattice yields, at total filling \(2/3\), a balanced fractional quantum
spin Hall (FQSH)-like state and a fully polarized \(2/3\) fractional Chern
insulator. Their competition separates two aspects of fractional-state
control: the finite-momentum intercomponent coupling governs the stability of
the balanced FQSH-like state, whereas a uniform pseudospin anisotropy shifts
the relative energies of conserved sectors and can switch the global ground
state between the balanced and polarized states. These results establish a
minimal lattice setting that connects topological-miniband reconstruction,
fractional-state formation, and fractional-state selection, illustrating how
distinct levels of control can be combined to navigate competing fractional
topological states.
\end{abstract}

\maketitle

\section{Introduction}
\label{sec:introduction}

The interplay of topology and strong electronic correlations provides a route
to quantum phases beyond the classification of noninteracting topological
matter~\cite{HasanKane2010,QiZhang2011}. A paradigmatic example is the
fractional Chern insulator (FCI), in which interactions stabilize fractional
quantum Hall--type topological order in a partially filled lattice Chern band
without requiring Landau
levels~\cite{TangMeiWen2011,SunGuKatsuraDasSarma2011,NeupertEtAl2011FCI,RegnaultBernevig2011,ShengGuSunSheng2011}.
Experiments have progressively brought such physics into solid-state
platforms, including graphene-based Hofstadter Chern
bands~\cite{SpantonEtAl2018} and zero-field moir\'e Chern
bands~\cite{CaiEtAl2023,ZengEtAl2023MoireFCI,ParkEtAl2023,XuEtAl2023,LuEtAl2024GrapheneFQAH,ParkEtAl2026DissipationlessFCI}.
A nonzero Chern number, however, is only the starting point for
fractionalization. Generic lattice Chern bands retain kinetic dispersion,
nonuniform Berry curvature and quantum metric, and Bloch-state form factors
that enter the projected
interaction~\cite{ParameswaranRoySondhi2012,Roy2014,JacksonMollerRoy2015,LedwithEtAl2020,WangEtAl2021Geometry}.
Fractional order therefore depends on the combined structure of topology, band
isolation, dispersion, quantum geometry, and
interactions~\cite{BergholtzLiu2013,ParameswaranRoySondhi2013,LiuBergholtz2024Review}. From this
perspective, engineering the underlying topological band is the first stage of
the many-body problem of creating fractional order.

Long-period scalar or electrostatic superlattices provide a direct real-space
route to reshaping these ingredients~\cite{ForsytheEtAl2018PatternedSuperlattice}.
Band folding and hybridization can
reconstruct both the dispersion and the Bloch wave functions of a parent
topological band while the superlattice period and amplitude remain externally
accessible parameters~\cite{GhorashiCano2023Multilayer}. Recent work has used this
strategy to generate narrow
topological minibands with favorable geometry and to realize fractional Chern
or fractional quantum anomalous Hall states in engineered
bands~\cite{GhorashiEtAl2023Superlattice,TanReddyFuDevakul2024}.
Patterned-gate studies have established superlattice routes to
Bernevig-Hughes-Zhang (BHZ)-type topological minibands~\cite{MiaoRashidiDai2025IdealBHZ}, while band-folding
analyses provide a broader route to engineering miniband
topology~\cite{YangLiuSchindlerLiu2025BandFolding}. Taken together, these
developments show that real-space reconstruction can bring conventional
lattice topological models into regimes favorable for fractionalization. In
multicomponent systems, the same strategy opens a richer many-body landscape
in which distinct fractional organizations can compete.

Time-reversal-related Chern sectors therefore enlarge the fractionalization
problem from state formation to competition among distinct many-body
organizations. Balanced occupation can support time-reversal-paired fractional
topological order, including fractional quantum spin Hall (FQSH)
states~\cite{LevinStern2009,Qi2011,NeupertEtAl2011FTI,LiShengTingChen2014,RepellinBernevigRegnault2014,WuShafferWuSantos2024FQSH},
while component polarization provides a competing route to chiral fractional
Hall order. Recent studies of opposite-Chern and moir\'e systems have shown
that balanced fractional topological states can lie close in energy to fully
or partially polarized competitors and that their stability is sensitive to
interaction structure and
screening~\cite{BrunnerNeupertWagner2026,WangEtAl2026FTICandidate,KwanEtAl2026FTI}.
Experimental reports of fractional quantum spin Hall behavior in
moir\'e MoTe$_2$~\cite{KangEtAl2024FQSH} and electric-field-controlled
competition between fractional Chern insulating and magnetic
states~\cite{ChangEtAl2026CompetingStates} make these questions
particularly timely.
This suggests two distinct control problems. One is to stabilize fractional
correlations within a given component sector; the other is to select
energetically among competing sectors. How these levels of control emerge
together from an engineered topological lattice band is the question we
address below.

The Qi-Wu-Zhang (QWZ) and BHZ models provide a minimal setting in which these questions can
be separated cleanly. The former provides a canonical lattice Chern insulator,
while the latter combines time-reversed blocks with opposite Chern
numbers~\cite{QiWuZhang2006,BernevigHughesZhang2006}. We use the same
scalar-superlattice construction to address three connected questions. Can a
physically reconstructed QWZ Chern miniband support fractional topological
order once interactions are treated explicitly? What fractional organizations
emerge when the time-reversed BHZ partner is restored? Finally, what controls
the stability of these states and which conserved sector is selected
energetically?

We find that the scalar superlattice reconstructs the QWZ band into an
isolated \(C=+1\) miniband with substantially improved Berry-curvature and
quantum-metric uniformity, and exact diagonalization identifies a \(1/3\) FCI
in the resulting projected problem. Restoring the time-reversed BHZ partner
reveals two competing fractional organizations at total filling \(2/3\), a
balanced FQSH-like state at weak intercomponent coupling and a fully polarized
\(2/3\) FCI. The finite-momentum intercomponent coupling controls the
stability of the balanced fractional state, whereas a uniform pseudospin
anisotropy changes the relative energetic ordering of conserved sectors and
can switch the global ground state between the balanced and polarized states.
Together, these results establish a hierarchy from band engineering to
fractional-state formation and state selection, suggesting a design principle
in which band structure, interaction-driven correlations, and component
energetics provide distinct control layers for competing fractional
topological states.
%
%
\section{Superlattice engineering of topological minibands}
\label{sec:band_engineering}

The central many-body question is whether a local scalar modulation can
reshape a canonical Chern band into an isolated platform for fractional
topological order. Nontrivial topology alone is not sufficient.
Within an isolated-band
description, a small residual bandwidth favors an interaction-dominated
regime, while separation from neighboring bands makes the single-band
description itself meaningful~\cite{BergholtzLiu2013}. At the same time,
the Bloch wave functions determine the projected density algebra and
interaction form factors, bringing Berry curvature and quantum geometry
directly into the interacting problem~\cite{ParameswaranRoySondhi2012,Roy2014,JacksonMollerRoy2015}.
We therefore first examine how the superlattice reconstructs both the
energetics and the Bloch-state geometry of the parent Chern band, and then
formulate the corresponding projected interacting problem.

\subsection{Parent topological bands and scalar superlattice}
\label{subsec:model_superlattice}

We set the primitive-lattice constant and the QWZ energy scale to unity.
With $\mathbf{k}=(k_x,k_y)$ and $\sigma_{x,y,z}$ denoting Pauli matrices in
the two-orbital space, the QWZ block is
\begin{equation}
\begin{aligned}
h_{\uparrow}(\mathbf{k})
={}&
\sin k_x\,\sigma_x+\sin k_y\,\sigma_y \\
&+
\left(m+\cos k_x+\cos k_y\right)\sigma_z,
\end{aligned}
\label{eq:qwz}
\end{equation}
where $m$ is the QWZ mass parameter. For $m=-1.3$, the positive-energy band
has Chern number $C=+1$ in our convention and serves as the parent of the
target superlattice miniband~\cite{QiWuZhang2006}. The spin-conserving BHZ
model is obtained by adjoining its time-reversed block,
\begin{equation}
h_{\mathrm{BHZ}}(\mathbf{k})
=
\begin{pmatrix}
h_{\uparrow}(\mathbf{k}) & 0\\
0 & h_{\downarrow}(\mathbf{k})
\end{pmatrix},
\qquad
h_{\downarrow}(\mathbf{k})
=
h_{\uparrow}^{*}(-\mathbf{k}),
\label{eq:bhz}
\end{equation}
so that the two conserved sectors have identical spectra and opposite Chern
numbers, $(C_{\uparrow},C_{\downarrow})=(+1,-1)$. Here $\uparrow$ and
$\downarrow$ label the two conserved, time-reversal-related sectors of the
spin-conserving BHZ model. We use ``pseudospin'' as a generic label, which may
correspond to the physical electron spin or another conserved internal
component~\cite{BernevigHughesZhang2006}.

For a square superlattice, the shortest reciprocal vectors form a
$C_4$-related star with magnitude $2\pi/a_M$. Retaining these four harmonics
with equal real Fourier amplitudes gives the minimal scalar modulation
\begin{equation}
V_M(\mathbf r)
=
2V_0
\left[
\cos\left(\frac{2\pi x}{a_M}\right)
+
\cos\left(\frac{2\pi y}{a_M}\right)
\right],
\label{eq:superlattice}
\end{equation}
where $\mathbf r=(x,y)$, $a_M$ is the superlattice period in units of the
primitive-lattice spacing, and $V_0$ is the amplitude of the leading Fourier
harmonic in the QWZ energy unit. Such lowest-harmonic scalar potentials
provide a minimal description of patterned-gate and artificial-superlattice
reconstruction of topological bands~\cite{MiaoRashidiDai2025IdealBHZ}. Because $V_M$ multiplies the identity in
orbital space and acts identically on the two time-reversed BHZ sectors, it
reconstructs their common spectrum without explicitly selecting one
chirality.

We fix $a_M=8$ and scan $V_0$, focusing below on $V_0=0.85$, for which
the folded spectrum contains the narrow, isolated $C=+1$ miniband highlighted
in Fig.~\ref{fig:band_engineering}. The enlarged real-space unit cell folds the
parent bands into the mini-Brillouin zone, while the scalar modulation
hybridizes the folded crossings and redistributes the parent-band spectral
weight among the resulting minibands~\cite{YangLiuSchindlerLiu2025BandFolding}. Additional single-particle
diagnostics and the complete miniband survey are summarized in
Appendix~\ref{app:miniband_geometry}.

\subsection{Energetic and geometric reconstruction of the \(C=+1\) band}
\label{subsec:miniband_geometry}

\begin{figure}[!t]
\centering
\includegraphics[width=\columnwidth]{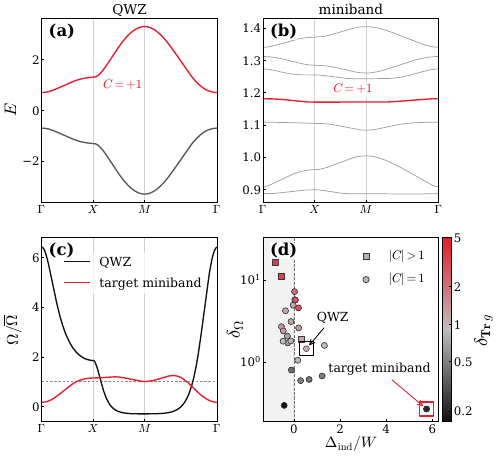}
\caption{\label{fig:band_engineering}
Superlattice reconstruction and target-band selection.
(a) QWZ parent dispersion along $\Gamma-X-M-\Gamma$, with the
positive-energy $C=+1$ band highlighted.
(b) Superlattice spectrum for $a_M=8$ and $V_0=0.85$ along the
corresponding mini-Brillouin-zone path. The highlighted $C=+1$ target
miniband has bandwidth $W=0.01075$ and
$\Delta_{\mathrm{ind}}/W=5.75$.
(c) Normalized Berry curvature $\Omega(\mathbf k)/\overline{\Omega}$ for
the QWZ parent and target miniband along the corresponding high-symmetry
paths. The horizontal dashed line marks the uniform-curvature value
$\Omega/\overline{\Omega}=1$.
(d) Full-Brillouin-zone Berry-curvature fluctuation $\delta_\Omega$
versus $\Delta_{\mathrm{ind}}/W$ for the topological minibands. Circles and
squares denote $|C|=1$ and $|C|>1$ minibands, respectively, while color
encodes the full-Brillouin-zone metric fluctuation
$\delta_{\mathrm{Tr}\,g}$. The vertical dashed line marks $\Delta_{\mathrm{ind}}=0$; arrows identify the
QWZ parent and target miniband.}
\end{figure}

The scalar superlattice reconstructs the parent QWZ band in two complementary
ways. It produces a pronounced hierarchy of one-body energy scales and, at
the same time, substantially homogenizes the Bloch-state geometry. The
positive-energy parent QWZ band is strongly dispersive
[Fig.~\ref{fig:band_engineering}(a)], whereas folding and hybridization
generate a narrow $C=+1$ miniband separated from the neighboring minibands
[Fig.~\ref{fig:band_engineering}(b)].

%
%
We characterize the target band energetically by its bandwidth $W$, the
energy variation across the Brillouin zone, and by the indirect gap
$\Delta_{\mathrm{ind}}$, defined as the separation between its complete
energy range and those of the neighboring bands. A positive
$\Delta_{\mathrm{ind}}$ therefore means that the target energy window does
not overlap neighboring bands anywhere in the Brillouin zone. For the
selected miniband, $\Delta_{\mathrm{ind}}/W\simeq5.75$, so the target band is narrow
compared with its global separation from neighboring minibands.

The reconstruction is not limited to the energy dispersion. We denote the
single-particle Berry curvature of the isolated band by
$\Omega(\mathbf k)$, with
$C=(2\pi)^{-1}\int_{\mathrm{BZ}}d^2k\,\Omega(\mathbf k)$.
Because the primitive Brillouin zone and mini-Brillouin zone have different
areas, their raw curvature magnitudes are not directly comparable.
Figure~\ref{fig:band_engineering}(c) therefore shows
$\Omega(\mathbf k)/\overline{\Omega}$ along the corresponding high-symmetry
paths, where $\overline{\Omega}$ denotes the full-Brillouin-zone average for
the band under consideration. This normalization removes the trivial change
of the mean curvature associated with the smaller mini-Brillouin zone while
retaining the momentum dependence of the curvature. Uniform Berry curvature
corresponds to $\Omega/\overline{\Omega}=1$. The QWZ parent varies strongly
along the path and changes sign, whereas the target miniband remains positive
and much closer to its mean value.

To quantify the geometric reconstruction over the full Brillouin zone,
rather than only along the displayed cuts, we use
$\delta_X\equiv\sigma_X/|\overline{X}|$ for
$X=\Omega$ or $X=\mathrm{Tr}\,g$, where $\overline{X}$ and $\sigma_X$ are
the full-Brillouin-zone mean and standard deviation, respectively. The
Berry-curvature fluctuation decreases from
$\delta_\Omega\simeq1.47$ in the QWZ parent to approximately $0.27$ in the
target miniband. The quantum metric $g_{\mu\nu}(\mathbf k)$ shows the same
trend: the relative fluctuation of its trace,
$\mathrm{Tr}\,g=g_{xx}+g_{yy}$, decreases from
$\delta_{\mathrm{Tr}\,g}\simeq1.02$ to approximately $0.21$. These
dimensionless measures therefore show that the superlattice substantially
homogenizes both geometric fields. Detailed definitions, full momentum-space
maps, and additional quantum-geometric diagnostics are collected in
Appendix~\ref{app:miniband_geometry}.

Figure~\ref{fig:band_engineering}(d) places the target within the broader set
of topological minibands generated at the same superlattice parameters by
plotting $\delta_\Omega$ against the signed ratio
$\Delta_{\mathrm{ind}}/W$. Positive values correspond to globally separated
energy windows, whereas negative values signal indirect overlap. Marker
shape distinguishes the Chern-number classes, while color encodes
$\delta_{\mathrm{Tr}\,g}$. The selected $C=+1$ miniband lies in the favorable lower-right part of this
energetic--geometric landscape, combining a large positive
$\Delta_{\mathrm{ind}}/W$ with comparatively small curvature and metric
fluctuations.

Taken together, Fig.~\ref{fig:band_engineering} shows that the scalar
superlattice does more than narrow the parent dispersion. The target
miniband combines global spectral isolation with substantially more uniform
quantum geometry. Such one-body characteristics are favorable for an
interacting Chern-band problem~\cite{BergholtzLiu2013,Roy2014,JacksonMollerRoy2015}, yet the emergence
of fractional topological order is ultimately a many-body question.

\subsection{Projected interacting problem}
\label{subsec:projected_model}

Having identified the target miniband, we project both its residual
dispersion and a two-dimensional Coulomb interaction onto this band.
Let $\gamma_{\sigma,\mathbf k}^{\dagger}$ create a particle in the target
miniband and $F_\sigma(\mathbf k,\mathbf q)$ denote its projected density form
factor. For momentum transfer $\mathbf q$, the projected density is
\begin{equation}
\bar{\rho}_{\sigma}(\mathbf q)
=
\sum_{\mathbf k}
F_{\sigma}(\mathbf k,\mathbf q)
\gamma_{\sigma,[\mathbf k+\mathbf q]_{\rm mBZ}}^{\dagger}
\gamma_{\sigma,\mathbf k}.
\label{eq:projected_density}
\end{equation}
The form factor retains the Bloch-state information entering the projected
interaction~\cite{ParameswaranRoySondhi2012}. The finite-momentum sewing
convention when $\mathbf k+\mathbf q$ crosses a mini-Brillouin-zone boundary
is given in Appendix~\ref{app:projected_finite_torus}.
The same superlattice reconstruction that narrows and isolates the
miniband therefore enters the interacting problem through the residual
dispersion $\varepsilon_{\sigma}(\mathbf k)$ and the momentum-dependent
form factors $F_{\sigma}(\mathbf k,\mathbf q)$.

We retain both in the effective single-miniband Hamiltonian,
\begin{equation}
\begin{aligned}
H_{\mathrm{proj}}
={}&
\sum_{\sigma,\mathbf k}
\varepsilon_{\sigma}(\mathbf k)
\gamma_{\sigma,\mathbf k}^{\dagger}
\gamma_{\sigma,\mathbf k}
\\
&+
\frac{1}{2N_{\phi}}
\sum_{\mathbf q\neq0}
\sum_{\sigma,\sigma'}
V_{\sigma\sigma'}(\mathbf q)
:\!\bar{\rho}_{\sigma}(-\mathbf q)
\bar{\rho}_{\sigma'}(\mathbf q)\!:,
\end{aligned}
\label{eq:projected_hamiltonian}
\end{equation}
where $N_{\phi}$ is the number of target-band orbitals per component and
\(:\!\cdots\!:\) denotes normal ordering. In Eq.~\eqref{eq:projected_hamiltonian},
the $\mathbf q=0$ transfer is omitted, with the uniform total-density Hartree
contribution canceled by the neutralizing background.

For the interaction we use the two-dimensional Coulomb kernel
$1/|\mathbf q|$ in momentum space. We absorb the overall intracomponent
prefactor into the interaction energy unit and parameterize the relative
intercomponent Coulomb strength by $\lambda$,
\begin{equation}
V_{\uparrow\uparrow}(\mathbf q)
=
V_{\downarrow\downarrow}(\mathbf q)
=
\frac{1}{|\mathbf q|},
\qquad
V_{\uparrow\downarrow}(\mathbf q)
=
V_{\downarrow\uparrow}(\mathbf q)
=
\frac{\lambda}{|\mathbf q|}.
\label{eq:interaction}
\end{equation}
Thus $\lambda=0$ describes decoupled time-reversed components, whereas
$\lambda=1$ gives equal bare intra- and intercomponent Coulomb amplitudes.
The QWZ calculation retains a single Chern component, while the BHZ problem
retains both time-reversed components and allows their relative interaction
strength to be varied.

Equation~\eqref{eq:projected_hamiltonian} defines an effective isolated-band
model and does not include interaction-induced mixing with neighboring
minibands. Momentum discretization, reciprocal-space conventions, and the
remaining projection details are collected in
Appendix~\ref{app:projected_finite_torus}. We next determine whether the
engineered $C=+1$ QWZ miniband develops fractional topological order.
\section{Fractional Chern order in the engineered QWZ miniband}
\label{sec:qwz_fci}

We now test the engineered $C=+1$ miniband at one-third filling. We study the
projected Hamiltonian by exact diagonalization in translation-resolved
momentum sectors. We consider $N=10$ particles in $N_\phi=30$ target-band
orbitals on a $6\times5$ torus, so that the band filling is
$\nu=N/N_\phi=1/3$. Both the residual miniband dispersion and the projected
Coulomb interaction introduced in Sec.~\ref{sec:band_engineering} are
retained. We denote the boundary twists along the two torus directions by
$\boldsymbol{\theta}=(\theta_x,\theta_y)$. For a Laughlin-type fractional
Chern insulator in a $C=+1$ band, one expects a threefold torus manifold
together with fractional Hall response and characteristic quasihole counting~\cite{WenNiu1990,RegnaultBernevig2011,WuBernevigRegnault2012}.

\subsection{Many-body manifold and topological response}
\label{subsec:qwz_topology}

\begin{figure}[t]
    \centering
    \includegraphics[width=\columnwidth]{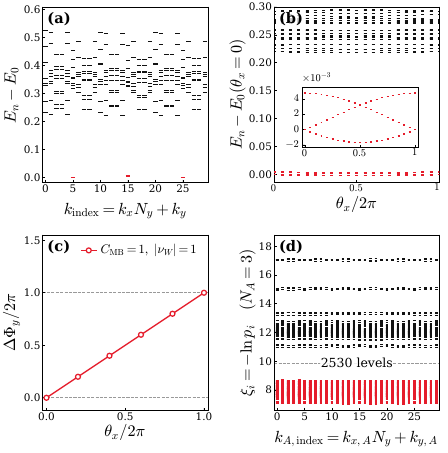}
    \caption{\label{fig:qwz_fci}
    Many-body identification of fractional Chern order in the engineered QWZ
    miniband on the $6\times5$ torus with $N=10$ particles in $N_\phi=30$
    target-band orbitals at $\nu=1/3$. (a) Zero-twist momentum-resolved many-body
    spectrum, with the lowest $D=3$ manifold highlighted. Momentum sectors are
    displayed using the flattened index $k_{\rm index}=k_xN_y+k_y$, where $k_x$
    and $k_y$ label the integer momentum-sector coordinates. (b) Spectral flow as
    $\theta_x$ is varied through one flux cycle at fixed $\theta_y=0$, with
    energies referenced to the zero-twist ground-state energy. The inset resolves
    the three lowest levels on a $10^{-3}$ energy scale, using the same twist
    interval and energy reference. (c) Unwrapped
    determinant Wilson-loop phase change of the isolated rank-three manifold
    relative to its value at \(\theta_x=0\), shown in units of \(2\pi\).
    Each transverse loop is taken over a complete \(\theta_y\) cycle. Its unit
    winding magnitude is \(|\nu_W|=1\); the same rank-three bundle has
    \(C_{\rm MB}=1\). (d) Momentum-resolved
    particle-entanglement spectrum of the equal-weight mixture of the three
    low-energy states at zero twist for the particle cut $N_A=3$; subsystem
    momenta use the analogous flattened index. The highlighted low-lying branch
    contains the predetermined 2530 Laughlin quasihole levels.}
\end{figure}

At zero boundary twist, the momentum-resolved spectrum in
Fig.~\ref{fig:qwz_fci}(a) displays a separated group of three lowest states,
consistent with the threefold torus structure of the $\nu=1/3$ Laughlin
state. Figure~\ref{fig:qwz_fci}(b) follows this low-energy manifold as
$\theta_x$ is varied through one flux cycle at fixed $\theta_y=0$; the
three states evolve as a separated group throughout the displayed flow,
providing the characteristic spectral response expected for a fractional
Chern state~\cite{RegnaultBernevig2011,ShengGuSunSheng2011,
WuBernevigRegnault2012}.
A full two-dimensional twist-space calculation further confirms that the
rank-three manifold remains isolated and numerically well defined over the
sampled twist torus; the corresponding gap and overlap checks are summarized
in Appendix~\ref{app:twists_chern_wilson}.

Because the low-energy states form a quasi-degenerate manifold, we
characterize their collective holonomy through the determinant Wilson-loop
phase of the rank-three subspace. Figure~\ref{fig:qwz_fci}(c) shows its
continuous evolution as $\theta_x$ advances through one period, with the
transverse loop taken over a complete $\theta_y$ cycle. The phase winds once
in magnitude, $|\nu_W|=1$, providing a direct visualization of the nontrivial
topology of the three-state manifold.

The same twist-space topology is quantified by the first Chern number of the
many-body bundle,
\begin{equation}
C_{\rm MB}
=
\frac{1}{2\pi}
\int_{0}^{2\pi}d\theta_x
\int_{0}^{2\pi}d\theta_y\,
\mathrm{Tr}\,
\mathcal F_{\theta_x\theta_y},
\label{eq:many_body_chern}
\end{equation}
where $\mathcal F_{\theta_x\theta_y}$ is the non-Abelian Berry curvature of
the rank-three manifold. We obtain $C_{\rm MB}=1$~\cite{NiuThoulessWu1985,Hatsugai2005Multiplet,FukuiHatsugaiSuzuki2005},
corresponding to the expected fractional Hall response $C_{\rm MB}/3=1/3$.
Details of the twist-space construction, determinant Wilson loop, and
bundle-validation conventions are given in
Appendix~\ref{app:twists_chern_wilson}.

\subsection{Entanglement structure and FCI identification}
\label{subsec:qwz_entanglement}

We next ask whether the same topological manifold carries the internal
correlations of the Laughlin state. We evaluate the particle-entanglement
spectrum (PES) of the equal-weight three-state mixture at zero twist using
a particle cut \(N_A=3\). The entanglement levels are \(\xi_i=-\ln p_i\),
where \(p_i\) are the eigenvalues of the particle reduced density matrix.
The construction and quasihole-counting conventions are summarized in
Appendix~\ref{app:pes_counting}. Translation symmetry further resolves
the spectrum into subsystem-momentum sectors.

The particle cut leaves fewer particles in the same orbital space,
making the Laughlin quasihole counting the relevant reference for the
low-lying entanglement spectrum~\cite{RegnaultBernevig2011,SterdyniakEtAl2011}. For a fermionic
\(\nu=1/3\) Laughlin state, the \((1,3)\) rule counts admissible orbital
configurations with at most one particle in any three consecutive
orbitals, including across the periodic boundary. With \(N_\phi=30\)
and \(N_A=3\), this gives 2530 states. Figure~\ref{fig:qwz_fci}(d) shows
precisely this number of levels in a branch separated from the remaining
entanglement spectrum. This agreement reveals the Laughlin-type
internal correlations of the three-state manifold whose topology was
established above.

Together with the spectral and twist-space results, this entanglement
structure identifies the ground-state manifold on the \(6\times5\)
torus as a \(\nu=1/3\) FCI in the superlattice-engineered QWZ miniband.
It provides the single-component fractional parent for the
time-reversal-paired problem considered next.
%
\section{Fractional states of the time-reversed BHZ minibands}
\label{sec:bhz}

Restoring the time-reversed BHZ partner introduces a second conserved Chern
sector and, with it, a new degree of freedom in how the particles organize. At
fixed total filling, particles can be distributed between two sectors of
opposite chirality in different proportions. We define the total filling,
normalized to the number \(N_\phi\) of target-band orbitals in one component,
as \(\nu_T=(N_\uparrow+N_\downarrow)/N_\phi=2/3\). Two limiting sector
configurations are especially simple. The balanced sector has
\(\nu_\uparrow=\nu_\downarrow=1/3\), whereas in a fully polarized sector one
component is empty and the other is filled to \(2/3\). We first determine the
many-body order realized in these two limiting sectors and then compare their
energetics when the component imbalance is allowed to vary.

\subsection{Balanced time-reversal-paired fractional state}
\label{subsec:bhz_balanced}

We begin with the balanced sector on a \(3\times5\) torus,
\(N_\uparrow=N_\downarrow=5\). In the decoupled limit \(\lambda=0\),
each component is a \(1/3\)-filled copy of the FCI established in
Sec.~\ref{sec:qwz_fci}, with the two components carrying opposite Chern
chirality. Their threefold torus manifolds therefore combine into the
natural nine-state parent of an FQSH state~\cite{LevinStern2009,NeupertEtAl2011FTI,LiShengTingChen2014,RepellinBernevigRegnault2014,WuShafferWuSantos2024FQSH}.
The relevant question is whether this paired fractional structure
survives once the two components interact.

%
%
\begin{figure}[!t]
\centering
\includegraphics[width=\columnwidth]{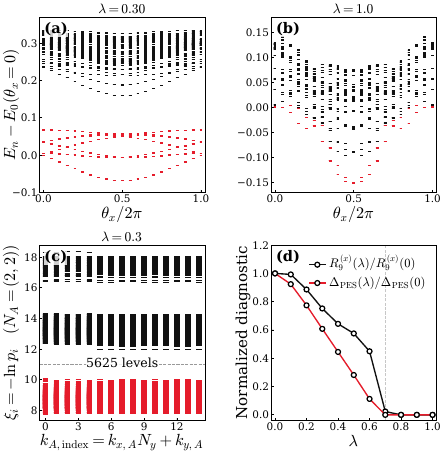}
\caption{
Interaction-driven evolution of the balanced FQSH-like state on the
\(3\times5\) torus with
\((N_\uparrow,N_\downarrow)=(5,5)\).
(a) Charge-flux spectral flow at \(\lambda=0.3\) as \(\theta_x\)
is varied at fixed \(\theta_y=0\).
Red levels denote the lowest nine states, which form an isolated
rank-nine manifold carrying charge and mixed spin-charge Chern numbers
\((C_c,C_{sc})=(0,6)\).
(b) Corresponding spectral flow at \(\lambda=1\), where the nine lowest
states merge with higher excitations. Energies in (a) and (b) are
referenced to the zero-twist ground-state energy.
(c) Momentum-resolved particle-entanglement spectrum at \(\lambda=0.3\)
for the equal mixture of the nine low-energy states and the cut
\((N_{A,\uparrow},N_{A,\downarrow})=(2,2)\). Subsystem momenta use
the flattened index \(k_{A,\mathrm{index}}=k_{x,A}N_y+k_{y,A}\).
Red levels form the predetermined 5625-level product-Laughlin branch.
(d) Normalized spectral-isolation ratio \(R_9^{(x)}\) and product-Laughlin
entanglement gap as functions of \(\lambda\), each normalized to its
\(\lambda=0\) value. Here \(R_9^{(x)}\) is the minimum separation above
the lowest nine states divided by their maximum internal width along the
displayed \(x\)-flux cycle. The dashed line at
\(\lambda=0.7\) marks the first sampled coupling at which the
full twist-space bundle-validity criterion
fails; it is not a thermodynamic phase boundary.
}
\label{fig:bhz35}
\end{figure}
For \(\lambda=0.3\), the lowest nine states remain separated from higher
excitations throughout the displayed charge-flux cycle
[Fig.~\ref{fig:bhz35}(a)]. The full two-dimensional twist-space
calculation further shows that the rank-nine manifold remains isolated
and numerically well defined over the sampled twist torus and carries
\((C_c,C_{sc})=(0,6)\), where \(C_c\) and \(C_{sc}\) are the many-body
Chern numbers obtained from charge and mixed spin-charge twists,
respectively. The vanishing charge response and nonzero mixed
spin-charge response distinguish this paired fractional state from a
chiral charge-Hall state. The complete twist conventions are given in
Appendix~\ref{app:twists_chern_wilson}. The particle-entanglement spectrum
provides an independent characterization of its internal correlations.
For the equal mixture of the nine states and the cut
\((N_{A,\uparrow},N_{A,\downarrow})=(2,2)\),
Fig.~\ref{fig:bhz35}(c) exhibits a clearly separated low-lying branch
containing the predetermined \(75^2=5625\) product-Laughlin quasihole
levels. The spectral, topological, and entanglement signatures therefore
consistently identify a finite-coupling FQSH-like state.

This structure is lost as the intercomponent interaction becomes strong.
At \(\lambda=1\), the contrast with the weak-coupling regime is visible
directly in Fig.~\ref{fig:bhz35}(b), where the nine lowest states merge
with higher levels during flux insertion and no longer form an isolated
low-energy manifold. Figure~\ref{fig:bhz35}(d) follows this loss of
fractional structure by comparing the normalized nine-state
spectral-isolation ratio \(R_9^{(x)}(\lambda)/R_9^{(x)}(0)\) with the
normalized product-Laughlin entanglement gap
\(\Delta_{\rm PES}(\lambda)/\Delta_{\rm PES}(0)\). Here
\(\Delta_{\rm PES}\) is the gap above the predetermined 5625-level branch.
Both quantities decrease strongly and collapse over the same sampled
interaction range. At \(\lambda=0.7\), the spectral isolation along the
displayed flux cycle is already nearly lost and the fixed-counting PES
gap has collapsed; this is also the first sampled coupling at which
the full twist-space bundle validation fails.

The balanced fractional state therefore survives a finite intercomponent
interaction before losing both the isolated many-body manifold and the
characteristic product-Laughlin entanglement structure. Together, these
results delineate the sampled finite-size stability window of the
balanced FQSH-like state and leave the character of the ensuing
strong-coupling regime as an open question.
%
%
\subsection{Polarized \texorpdfstring{\(2/3\)}{2/3} fractional Chern state}
\label{subsec:bhz_polarized}

We next consider the fully polarized sectors,
\((N_\uparrow,N_\downarrow)=(N,0)\) and \((0,N)\). In either case the
intercomponent interaction vanishes identically, so \(\lambda\) drops
out and the problem reduces to a single superlattice-engineered QWZ
miniband at filling \(\nu=2/3\). This is the complementary filling of
the engineered Chern miniband studied at \(\nu=1/3\) in
Sec.~\ref{sec:qwz_fci}, and related polarized \(2/3\) fractional Chern
states have been established in moir\'e settings theoretically and
experimentally~\cite{WangEtAl2024MoTe2,ZengEtAl2023MoireFCI}.
The projected Hamiltonian considered here retains the miniband
dispersion and lattice form factors, and an exact particle--hole symmetry
is not assumed. We therefore characterize the \(2/3\) state directly
through its spectral, topological, and entanglement signatures.

%
%
\begin{figure}[!t]
\centering
\includegraphics[width=\columnwidth]{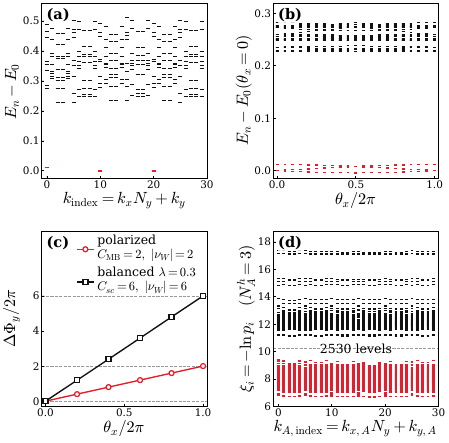}
\caption{
Polarized \(2/3\) fractional Chern state and comparison with the
balanced BHZ response.
(a) Zero-twist momentum-resolved spectrum on the \(6\times5\) torus
with \(N=20\) particles in \(N_\phi=30\) target-band orbitals.
Red levels denote the isolated rank-three low-energy manifold. Momentum
sectors use the flattened index \(k_{\rm index}=k_xN_y+k_y\).
(b) Charge-flux spectral flow of the same manifold as \(\theta_x\)
is varied through one flux cycle at fixed \(\theta_y=0\).
(c) Unwrapped determinant Wilson-loop phase changes relative to their
values at \(\theta_x=0\), shown in units of \(2\pi\), for the polarized
rank-three charge bundle, with \(C_{\rm MB}=2\) and \(|\nu_W|=2\), and
the balanced rank-nine mixed spin-charge bundle at \(\lambda=0.3\),
with \(C_{sc}=6\) and \(|\nu_W|=6\). Each transverse loop is taken over
a complete \(\theta_y\) cycle.
(d) Momentum-resolved three-hole particle-entanglement spectrum of
the polarized state for \(N_A^h=3\); hole-subsystem momenta use the
analogous flattened index. Red levels form the expected
2530-state \((1,3)\)-admissible Laughlin branch.
}
\label{fig:polarized23}
\end{figure}
For the displayed polarized state, we occupy the \(C=+1\) component and
consider \(N=20\) electrons in \(N_\phi=30\) target-band orbitals on a
\(6\times5\) torus. Figure~\ref{fig:polarized23}(a) shows a sharply
isolated three-state manifold at zero twist. Its isolation persists over
boundary twists, with the minimum gap above the manifold exceeding its
maximum internal width by more than an order of magnitude. The same
three states remain separated from higher excitations throughout the
charge-flux cycle in Fig.~\ref{fig:polarized23}(b). The associated
rank-three bundle carries \(C_{\rm MB}=2\), and the transverse-\(y\)
determinant Wilson phase winds twice in magnitude, \(|\nu_W|=2\), as
\(\theta_x\) spans one period. These spectral and topological signatures
identify a well-resolved \(2/3\) fractional Chern state.

Its internal correlations admit a particularly transparent hole
description. The 20-electron state in 30 orbitals contains \(N_h=10\)
holes, corresponding to a \(1/3\)-filled hole sector. For the equal
mixture of the three low-energy states, we retain \(N_A^h=3\) holes in
the particle-entanglement partition. Figure~\ref{fig:polarized23}(d)
displays a separated low-lying branch containing the expected 2530
\((1,3)\)-admissible Laughlin levels. The \(2/3\) electron state therefore
exhibits the correlations of a \(1/3\)-filled Laughlin-like hole fluid.

Figure~\ref{fig:polarized23}(c) places the nontrivial Wilson responses of
the two fractional organizations side by side. The polarized rank-three
charge bundle has \(C_{\rm MB}=2\), while the balanced rank-nine state
at \(\lambda=0.3\) carries \(C_{sc}=6\) in the mixed spin-charge channel
and has vanishing charge response. Under the adopted twist convention,
normalizing by the respective manifold ranks gives
\(C_{\rm MB}/3=C_{sc}/9=2/3\). The same normalized fractional response
thus appears in different channels, through the mixed spin-charge
response for the balanced state and directly through the charge response
for the polarized state. On the common \(3\times5\) cluster used for the
energetic comparison below, the polarized three-state manifold likewise
carries \(C_{\rm MB}=2\), as summarized in Table~\ref{tab:app_bundle_validity}.
%
\subsection{Sector competition and pseudospin anisotropy}
\label{subsec:bhz_competition}

Having established the balanced and polarized fractional states
separately, we next ask which one is selected energetically at fixed
total particle number. Related competition between balanced fractional
topological states and polarized \(2/3\) fractional Hall states has
been found in other opposite-Chern and moir\'e settings~\cite{BrunnerNeupertWagner2026,WangEtAl2026FTICandidate,KwanEtAl2026FTI}.
Here, the separate conservation of \(N_\uparrow\) and \(N_\downarrow\)
allows this competition to be resolved into distinct number sectors.
On the common \(3\times5\) cluster, we fix
\(N=N_\uparrow+N_\downarrow=10\) and label the sectors by the
polarization \(P=|N_\uparrow-N_\downarrow|/N\).

The finite-momentum intercomponent coupling \(\lambda\) changes both
correlations and sector energies. A uniform component-imbalance energy
provides an additional way to control their relative ordering.
We represent this contribution by the uniform pseudospin anisotropy
\(g_z(N_\uparrow-N_\downarrow)^2/(4N_\phi)\), added to \(H_{\rm proj}\).
Because this term is constant within each fixed
\((N_\uparrow,N_\downarrow)\) sector, it shifts the sector energies
without changing the corresponding eigenstates. Its microscopic
interpretation~\cite{JungwirthMacDonald2000} and relation to the
\(\mathbf q=0\) convention are discussed in
Appendix~\ref{app:sector_energetics}.

%
%
\begin{figure}[t]
\centering
\includegraphics[width=\columnwidth]{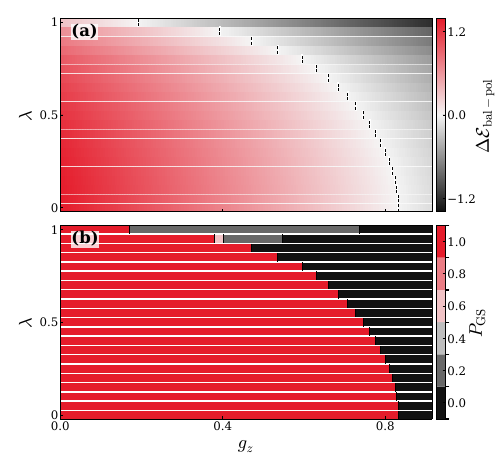}
\caption{
Energetic competition between balanced and polarized fractional states
on the \(3\times5\) torus.
(a) Energy difference
\(\Delta\mathcal E_{\rm bal-pol}
=\mathcal{E}_{P=0}-\mathcal{E}_{P=1}\)
as a function of the finite-\(q\) intercomponent coupling \(\lambda\)
and uniform pseudospin anisotropy \(g_z\).
Positive values indicate that the polarized sector is lower in energy,
while negative values favor the balanced sector.
Dashed segments mark
\(\Delta\mathcal E_{\rm bal-pol}=0\),
the balanced--polarized degeneracy.
(b) Ground-sector polarization \(P_{\rm GS}\) obtained by minimizing
over all six conserved polarization sectors. Partially polarized sectors
enter the lower envelope at the strongest sampled intercomponent couplings.
Horizontal white gaps separate independently calculated \(\lambda\)
values and carry no physical meaning.
}
\label{fig:sector_competition}
\end{figure}
To compare the sectors, let \(\mathcal E_P(\lambda,g_z)\) denote the
lowest energy at polarization \(P\), after minimization over the
many-body momentum sectors. Figure~\ref{fig:sector_competition}(a)
plots the balanced--polarized difference
\(\Delta\mathcal E_{\rm bal-pol}=\mathcal E_{P=0}-\mathcal E_{P=1}\) across the
\((\lambda,g_z)\) plane. At \(g_z=0\), this difference is positive for
every sampled \(\lambda\), so the polarized sector lies below the
balanced sector. Because the intercomponent interaction vanishes when
one component is empty, the polarized energy is independent of
\(\lambda\), whereas increasing \(\lambda\) lowers the balanced sector
relative to it. Positive \(g_z\) penalizes polarization and drives
\(\Delta\mathcal E_{\rm bal-pol}\) through zero, with the balanced--polarized
degeneracy moving to smaller \(g_z\) as \(\lambda\) increases.

This pairwise comparison determines when the two limiting sectors
exchange order, but another conserved sector may lie below both.
Figure~\ref{fig:sector_competition}(b) therefore addresses the stronger
question of which sector is globally lowest. For \(N=10\), the allowed
polarizations are \(P=0,0.2,\ldots,1\), and the ground-sector
polarization \(P_{\rm GS}\) is obtained by minimizing
\(\mathcal E_P(\lambda,g_z)\) over all six sectors. For the sampled
couplings through \(\lambda=0.9\), increasing \(g_z\) drives a direct
change from \(P=1\) to \(P=0\). At the two strongest sampled couplings,
partially polarized sectors enter the lower envelope. At
\(\lambda=0.95\), the sequence is
\(P=1\rightarrow0.6\rightarrow0.2\rightarrow0\), while at
\(\lambda=1\) the \(P=0.2\) sector intervenes between the polarized
and balanced sectors.

Where the balanced FQSH-like state and polarized \(2/3\) FCI both retain
their characteristic diagnostics, tuning \(g_z\) can therefore switch
the global ground sector between them without changing their internal
correlations. The partially polarized sectors appear at stronger
intercomponent coupling, beyond the sampled stability window of the
balanced FQSH-like state identified in Sec.~\ref{subsec:bhz_balanced}.
The energetic competition in this regime thus extends beyond the two
characterized fractional states.
\section{Discussion and conclusion}
\label{sec:conclusion}

Our results separate three levels of control in a multicomponent
fractional topological system. The scalar superlattice first reconstructs
the QWZ band into an isolated, narrow Chern miniband with substantially
improved quantum geometry, in which interactions stabilize a \(1/3\)
FCI. Restoring the time-reversed BHZ partner enlarges this fractionalized
setting to include competing balanced and polarized organizations,
represented here by an FQSH-like state and a polarized \(2/3\) FCI.
Their competition shows that forming a correlated state within a
conserved sector and selecting the global ground sector are distinct
problems. In the present model, the finite-momentum intercomponent
coupling \(\lambda\) changes both the correlated states and their
relative energies, whereas the uniform pseudospin anisotropy \(g_z\)
provides a way to change sector ordering without altering the eigenstates
within each sector.

Superlattice and electrostatic engineering have established routes to
reconstruct topological bands and create conditions favorable for
fractional Chern states~\cite{GhorashiEtAl2023Superlattice,TanReddyFuDevakul2024,MiaoRashidiDai2025IdealBHZ,YangLiuSchindlerLiu2025BandFolding},
while recent studies of opposite-Chern systems emphasize competition
among balanced, partially polarized, and fully polarized fractional
states~\cite{BrunnerNeupertWagner2026,WangEtAl2026FTICandidate,KwanEtAl2026FTI}.
The present results connect these perspectives within an engineered
lattice setting. Engineering a favorable topological miniband does not
by itself determine which correlated state becomes the ground state
once additional conserved components are present. A sector can support
fractional order without being energetically selected. Assessing the
stability of that order and its energetic competition with other
sectors are therefore complementary requirements, making polarization
energetics part of the design problem rather than a consequence of
band flattening alone.

Larger-system studies can determine how the balanced, polarized, and
partially polarized competitors evolve with size and clarify the nature
of the boundaries between them. At the microscopic level, connecting
the effective parameters \(\lambda\) and \(g_z\) to screening, gate
geometry, and component-dependent interactions would turn the controls
identified here into experimentally tunable knobs~\cite{KwanEtAl2026FTI,JungwirthMacDonald2000}.
These directions offer an opportunity to extend the present separation
of control tasks to concrete platforms. More broadly, applying this
perspective to spin-, valley-, layer-, and orbital-resolved Chern systems
could enable the engineering not only of conditions favorable for
fractionalization, but also of the energetic selection among distinct
fractional topological states. Treating band engineering, correlation
engineering, and state selection as distinct but cooperating tasks may
thus provide a systematic route to navigating competing fractional
phases in engineered Chern systems.
\begin{acknowledgments}
This research was supported by the National Natural Science Foundation of
China (Grant Nos.\ 12174167 and 12247101), the Fundamental Research Funds
for the Central Universities (Grant No.\ lzujbky-2025-jdzx07), and the
Natural Science Foundation of Gansu Province (No.\ 25JRRA799).
\end{acknowledgments}

\section*{Data Availability}
The data supporting this study will be made openly available in the final
version~\cite{zenodo_cite}.

\appendix
\makeatletter
\@addtoreset{figure}{section}
\makeatother
\renewcommand{\thefigure}{\Alph{section}\arabic{figure}}
\providecommand{\theHfigure}{}
\renewcommand{\theHfigure}{appendix.\Alph{section}.\arabic{figure}}
%
%
%
%

\section{Superlattice minibands and quantum geometry}
\label{app:miniband_geometry}

\subsection{Miniband isolation and topology}
\label{app:miniband_isolation}

\begin{figure}[!t]
\centering
\includegraphics[width=\columnwidth]{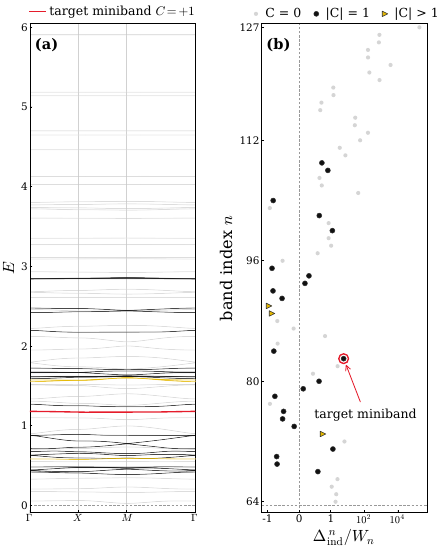}
\caption{\label{fig:app_full_miniband_context}
\textbf{Symmetry-reduced miniband spectrum and energetic--topological survey.}
(a) One half of the \(E\leftrightarrow-E\) symmetric miniband spectrum
of one QWZ block for \(m=-1.3\), \(a_M=8\), and \(V_0=0.85\),
plotted along \(\Gamma-X-M-\Gamma\). The horizontal dashed line marks
the \(E=0\) symmetry line. Minibands with \(|C|>1\) are shown in gold,
and the selected \(C=+1\) target miniband is highlighted in red.
(b) Signed ratio \(\Delta_{\mathrm{ind},n}/W_n\) for the same
half-spectrum, with \(n\) denoting the zero-based band index. The
vertical dashed line at \(\Delta_{\mathrm{ind},n}/W_n=0\) separates
globally isolated bands from bands with indirect energy-window
overlap. Bands are grouped as \(C=0\), \(|C|=1\), and \(|C|>1\);
the target \(n=83\) is marked separately. Higher-Chern bands are
included for spectral context.}
\end{figure}

At the representative superlattice period \(a_M=8\) used in
Sec.~\ref{sec:band_engineering}, we scan \(V_0\) and select
\(V_0=0.85\), which yields a narrow, globally isolated \(C=+1\)
miniband suitable for the projected many-body calculation.
Figure~\ref{fig:app_full_miniband_context} places the selected band
within the reconstructed spectrum and the energetic--topological survey.

For bands ordered by increasing energy, define
\(\varepsilon_n^{\min}=\min_{\mathbf k}\varepsilon_n(\mathbf k)\)
and
\(\varepsilon_n^{\max}=\max_{\mathbf k}\varepsilon_n(\mathbf k)\),
with extrema over the mini-Brillouin zone. The bandwidth is
\(W_n=\varepsilon_n^{\max}-\varepsilon_n^{\min}\), and the signed
indirect separation of an interior band is
\begin{equation}
\Delta_{\mathrm{ind},n}
=
\min\!\left\{
\varepsilon_{n+1}^{\min}-\varepsilon_n^{\max},
\,
\varepsilon_n^{\min}-\varepsilon_{n-1}^{\max}
\right\}.
\label{eq:app_indirect_gap}
\end{equation}
Positive \(\Delta_{\mathrm{ind},n}\) means that the complete energy
window of band \(n\) overlaps neither neighboring band. An
individual-band Chern number requires same-momentum nondegeneracy,
a weaker condition than positive indirect separation. The selected
band has zero-based index \(n=83\), bandwidth \(W=0.01075\), and
\(\Delta_{\mathrm{ind}}=0.06185\), giving
\(\Delta_{\mathrm{ind}}/W\simeq5.75\).

Before assigning a miniband Chern number, we check same-momentum
isolation on a \(41\times41\) mini-Brillouin-zone mesh, explicitly
including the relevant high-symmetry points and using an isolation
threshold of \(10^{-7}\). We then apply the Fukui--Hatsugai--Suzuki
method~\cite{FukuiHatsugaiSuzuki2005} using the embedding-aware
overlaps
\(\langle u_{\mathbf k}|e^{-i\delta\mathbf k\cdot\mathbf r}
|u_{\mathbf k+\delta\mathbf k}\rangle\)
for a mesh displacement \(\delta\mathbf k\). Here
\(|u_{\mathbf k}\rangle\) is the normalized cell-periodic Bloch state,
and \(\mathbf r\) retains the orbital positions within the supercell.
Only the selected \(C=+1\) miniband enters the interacting calculation
for this QWZ block.

\subsection{Full-Brillouin-zone quantum geometry}
\label{app:full_bz_geometry}

\begin{figure}[!t]
\centering
\includegraphics[width=\columnwidth]{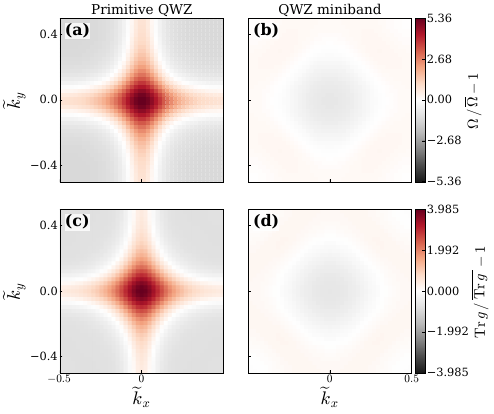}
\caption{\label{fig:app_full_bz_geometry}
\textbf{Full-Brillouin-zone quantum geometry.}
Normalized deviations
\(D_X(\mathbf k)\) for the primitive positive-energy QWZ band and
the selected superlattice miniband. Panels (a,b) show the
Berry-curvature deviations for the parent and target bands,
respectively, while panels (c,d) show the corresponding
quantum-metric-trace deviations. The plotted momentum coordinates
are \(\widetilde k_\mu=k_\mu/(2\pi)\) for the primitive band and
\(\widetilde k_\mu=a_Mk_\mu/(2\pi)\) for the target miniband.
Each row uses a common symmetric, zero-centered color scale,
so the parent and target fluctuations are compared without
independent panel-by-panel rescaling.}
\end{figure}

For the same Bloch states, with orbital embedding retained, we use
the Berry-connection convention
\(A_\mu=-i\langle u_{\mathbf k}|\partial_\mu u_{\mathbf k}\rangle\)
and curvature
\(\Omega=\partial_{k_x}A_y-\partial_{k_y}A_x\), where
\(\partial_\mu=\partial/\partial k_\mu\). The quantum metric is~\cite{Roy2014,JacksonMollerRoy2015}
\begin{equation}
g_{\mu\nu}(\mathbf k)
=
\operatorname{Re}
\langle\partial_\mu u_{\mathbf k}|
\bigl(1-|u_{\mathbf k}\rangle\langle u_{\mathbf k}|\bigr)
|\partial_\nu u_{\mathbf k}\rangle,
\label{eq:app_quantum_metric}
\end{equation}
with \(\operatorname{Tr}g=g_{xx}+g_{yy}\). The link-based numerical
implementation retains the orbital embedding and does not require
a globally smooth Bloch gauge.

We evaluate full-zone averages and population standard deviations
on \(41\times41\) equally weighted momentum grids in each band's
Brillouin zone. Using the mean \(\overline X\) and relative fluctuation
\(\delta_X\) defined in Sec.~\ref{subsec:miniband_geometry},
Fig.~\ref{fig:app_full_bz_geometry} displays the normalized deviations
\(D_X(\mathbf k)=[X(\mathbf k)-\overline X]/|\overline X|\)
for \(X=\Omega\) or \(X=\operatorname{Tr}g\). These deviations have
zero mean and variance \(\delta_X^2\), relating the full-zone maps
directly to the fluctuation measures used in the main text.

The full-zone maps complement the high-symmetry-path comparison in
Sec.~\ref{subsec:miniband_geometry}. On these grids, \(\delta_\Omega\)
decreases from \(1.4716\) for the primitive positive-energy QWZ band
to \(0.2708\) for the target miniband, while
\(\delta_{\operatorname{Tr}g}\) decreases from \(1.0201\) to \(0.2080\).
The parent curvature contains a small negative sampled region,
whereas the target curvature is positive at every sampled momentum.
\section{Folded-basis projection and finite-torus momentum conventions}
\label{app:projected_finite_torus}

\subsection{Folded-basis projection and momentum sewing}

To specify the form factor entering the projected density in
Eq.~\eqref{eq:projected_density}, we follow the standard projected-band
construction~\cite{ParameswaranRoySondhi2012} in the folded basis
\(\{|\mathbf k+\mathbf G,\alpha\rangle\}\), where
\(\mathbf G=2\pi(g_x,g_y)/a_M\) labels the folded momenta and
\(\alpha\) labels the two QWZ orbitals. For \(a_M=8\), there are
\(a_M^2=64\) folded momentum sectors and hence 128 basis states per
conserved component. We denote the target-band eigenvector in this
basis by \(\mathbf u_{\sigma,\mathbf k}\). Its supercell-site
representation \(\mathbf v_{\sigma,\mathbf k}\), used in the numerical
implementation, is related by the unitary supercell Fourier
transformation
\(\mathbf v_{\sigma,\mathbf k}=T(\mathbf k)\mathbf u_{\sigma,\mathbf k}\).

For a momentum transfer \(\mathbf q\), let
\(\mathbf k'=[\mathbf k+\mathbf q]_{\mathrm{mBZ}}\) denote the momentum
reduced to the mini-Brillouin zone. The target-band form factor is
\begin{equation}
\begin{aligned}
F_\sigma(\mathbf k,\mathbf q)
&=
\mathbf u^\dagger_{\sigma,\mathbf k'}
\mathcal S(\mathbf k,\mathbf q)
\mathbf u_{\sigma,\mathbf k},\\
\mathcal S(\mathbf k,\mathbf q)
&=
T^\dagger(\mathbf k')D(\mathbf q)T(\mathbf k),
\end{aligned}
\label{eq:app_form_factor_sewing}
\end{equation}
where \(D(\mathbf q)\) is diagonal in the supercell-site basis with
matrix element \(e^{i\mathbf q\cdot\mathbf R}\), and \(\mathbf R\)
denotes the site position within the supercell. In the adopted folded
plane-wave convention, \(\mathcal S\) identifies \(\mathbf G\) with
\(\mathbf G+\mathbf Q\) modulo the primitive reciprocal lattice, where
\(\mathbf Q=\mathbf k+\mathbf q-\mathbf k'\) is a superlattice
reciprocal vector. The site-basis density phase is therefore
incorporated into the sewing map. Both QWZ orbitals share the same
supercell position \(\mathbf R\), so no additional orbital-dependent
sewing phase is present. The construction is gauge covariant: a phase
redefinition of the target-band eigenvectors is compensated by the
corresponding band operators, leaving the projected density invariant.

\subsection{Finite-torus and normal-ordering conventions}

On an \(N_x\times N_y\) torus of supercells, \(N_\phi=N_xN_y\).
For the clusters studied here, we call the dimension divisible by
three the \(x\) direction, so the \(N_\phi=30\) and \(N_\phi=15\)
clusters are denoted \(6\times5\) and \(3\times5\), respectively.
At zero boundary twist, the target-band momenta and momentum transfers
are \(k_\mu=2\pi j_\mu/(N_\mu a_M)\) and
\(q_\mu=2\pi s_\mu/(N_\mu a_M)\), with \(\mu=x,y\) and
\(j_\mu,s_\mu=0,\ldots,N_\mu-1\). For evaluating \(|\mathbf q|\)
in the Coulomb kernel of Eq.~\eqref{eq:interaction}, each \(s_\mu\)
is replaced by its shortest periodic representative:
\(\widetilde s_\mu=s_\mu\) for \(s_\mu\leq\lfloor N_\mu/2\rfloor\),
and \(\widetilde s_\mu=s_\mu-N_\mu\) otherwise. The kernel is
evaluated only on these discrete nonzero transfers; no additional
reciprocal-lattice shell or finite-\(q\) regularization is included.

The colons in Eq.~\eqref{eq:projected_hamiltonian} denote normal
ordering with respect to the empty projected-band Fock vacuum.
For the form-factor convention in Eq.~\eqref{eq:app_form_factor_sewing},
\begin{equation}
\begin{aligned}
:\bar\rho_\sigma(-\mathbf q)\bar\rho_{\sigma'}(\mathbf q):
={}&
\bar\rho_\sigma(-\mathbf q)\bar\rho_{\sigma'}(\mathbf q)\\
&-
\delta_{\sigma\sigma'}
\sum_{\mathbf k}|F_\sigma(\mathbf k,\mathbf q)|^2
n_{\sigma,\mathbf k},
\end{aligned}
\label{eq:app_normal_ordering}
\end{equation}
where
\(n_{\sigma,\mathbf k}
=\gamma^\dagger_{\sigma,\mathbf k}\gamma_{\sigma,\mathbf k}\).
The numerical implementation constructs the corresponding quartic
operator directly. The subtracted same-component contraction is
generally a momentum-dependent one-body operator and is not absorbed
into the residual target-band dispersion. Normal ordering applies
to each retained nonzero-\(\mathbf q\) density product and is distinct
from omitting the \(\mathbf q=0\) transfer in
Eq.~\eqref{eq:projected_hamiltonian}.
%
%
%
%

\section{Many-body twist-space topology and bundle validation}
\label{app:twists_chern_wilson}

\subsection{Twist-space multiplet topology}

We follow the standard construction of Chern numbers for multiplets
under twisted boundary conditions and its lattice-gauge discretization~\cite{NiuThoulessWu1985,Hatsugai2005Multiplet,FukuiHatsugaiSuzuki2005}.
Extending the finite-torus conventions of
Appendix~\ref{app:projected_finite_torus}, boundary phases shift the
target-band momenta to
\(k_{\mu,\sigma}=(2\pi j_\mu+\theta_{\mu,\sigma})/(N_\mu a_M)\),
with \(\mu=x,y\) and \(j_\mu=0,\ldots,N_\mu-1\). The single-component
problem uses ordinary charge twists to define \(C_{\rm MB}\). In BHZ,
common twists
\(\boldsymbol\theta_\uparrow=\boldsymbol\theta_\downarrow
=(\theta_x,\theta_y)\) define \(C_c\), while mixed spin-charge twists
\(\boldsymbol\theta_\uparrow=(\theta_x,\theta_y)\) and
\(\boldsymbol\theta_\downarrow=(-\theta_x,\theta_y)\) define \(C_{sc}\).
In each case, \(\theta_x\) and \(\theta_y\) are independent coordinates
of the two-dimensional torus \(\boldsymbol\theta\in[0,2\pi)^2\).

%
%
%
\begin{table*}[t]
\renewcommand{\thetable}{\Alph{section}\arabic{table}}
\renewcommand{\theHtable}{appendix.\Alph{section}.\arabic{table}}
\caption{Representative full-torus validation results. All listed rows pass
the corresponding sampled-torus criteria. The signed transverse-\(y\)
winding \(\nu_W\) is obtained from the same determinant-link and
plaquette data as \(C\). The final row gives the size-matched
\(3\times5\) polarized check on a \(7\times7\) twist mesh.}
\label{tab:app_bundle_validity}
\begin{ruledtabular}
\begin{tabular}{lccccccc}
State / channel & \(D\) & Twist mesh & Invariant & \(\nu_W\) &
\(\Delta_D^{\min}\) & \(s_{\rm link}^{\min}\) &
\(\min|\det\mathbf M_\mu|\) \\
\hline
QWZ \(1/3\), native \(6\times5\) & 3 & \(5\times5\) &
\(C_{\rm MB}=1\) & 1 & 0.214755 & 0.931908 & 0.809557 \\
Balanced BHZ \(\lambda=0.3\), charge, \(3\times5\) & 9 & \(5\times5\) &
\(C_c=0\) & 0 & 0.107877 & 0.813967 & 0.164349 \\
Balanced BHZ \(\lambda=0.3\), mixed, \(3\times5\) & 9 & \(5\times5\) &
\(C_{sc}=6\) & 6 & 0.105484 & 0.815740 & 0.168179 \\
Polarized \(2/3\), native \(6\times5\) & 3 & \(5\times5\) &
\(C_{\rm MB}=2\) & 2 & 0.216764 & 0.867128 & 0.652220 \\
Polarized \(2/3\), \(3\times5\) same-size check & 3 & \(7\times7\) &
\(C_{\rm MB}=2\) & 2 & 0.123758 & 0.900282 & 0.734934 \\
\end{tabular}
\end{ruledtabular}
\end{table*}
At each twist, we evaluate the target-band dispersion and form factors
at the shifted momenta and diagonalize the projected Hamiltonian
independently. The lowest \(D\) states are selected globally across
the many-body momentum sectors, with energies ordered as
\(E_0\leq E_1\leq\cdots\). We use \(D=3\) for the QWZ \(1/3\) and
polarized \(2/3\) states, and \(D=9\) for the balanced BHZ state.
The multiplet is selected by energy at each twist, rather than by
tracking states between neighboring twists.

Overlaps between neighboring multiplets include the single-particle
overlaps of their twist-dependent target-band orbitals. Across a
\(2\pi\) boundary, the large-gauge identification shifts the discrete
momentum index in the wrapped direction. Charge twists shift each
occupied component by \(+1\); mixed \(x\) twists shift \(\uparrow\)
by \(+1\) and \(\downarrow\) by \(-1\), while mixed \(y\) twists shift
both by \(+1\). The orbital relabeling includes the fermionic sign
from reordering the mapped occupations. With this identification
included in the bra-ket, a mesh step \(\delta\boldsymbol\theta_\mu\)
defines
\begin{equation}
\begin{aligned}
{}[\mathbf M_\mu(\boldsymbol\theta)]_{ab}
&=
\langle\Psi_a(\boldsymbol\theta)|
\Psi_b(\boldsymbol\theta+\delta\boldsymbol\theta_\mu)\rangle,\\
U_\mu(\boldsymbol\theta)
&=
\frac{\det\mathbf M_\mu(\boldsymbol\theta)}
{|\det\mathbf M_\mu(\boldsymbol\theta)|},\\
C
&=
\frac{1}{2\pi}\sum_{\boldsymbol\theta}\arg\Bigl[
U_x(\boldsymbol\theta)
U_y(\boldsymbol\theta+\delta\boldsymbol\theta_x)\\
&\qquad\qquad\times
U_x^*(\boldsymbol\theta+\delta\boldsymbol\theta_y)
U_y^*(\boldsymbol\theta)\Bigr].
\end{aligned}
\label{eq:app_bundle_chern}
\end{equation}
Here \(a,b=0,\ldots,D-1\), and \(\arg\in(-\pi,\pi]\) uses the
plaquette orientation \(+x,+y,-x,-y\). The invariant \(C\) denotes
\(C_{\rm MB}\), \(C_c\), or \(C_{sc}\), according to the chosen twist
torus. The determinant links represent the \(U(1)\) part of the
non-Abelian bundle connection. Their oriented plaquette phases
therefore discretize \(\operatorname{Tr}\mathcal F_{\theta_x\theta_y}\),
relating Eq.~\eqref{eq:app_bundle_chern} to the continuum Chern-number
definition in Eq.~\eqref{eq:many_body_chern}.

For a transverse \(y\) loop, we use
\(\det\mathcal W_y(\theta_x)=\prod_j U_y(\theta_x,\theta_{y,j})\),
including the periodic closing link. To obtain the unwrapped phase
change displayed in Figs.~\ref{fig:qwz_fci}(c) and
\ref{fig:polarized23}(c), we sum the oriented plaquette fluxes in each
strip between neighboring \(\theta_x\) values and accumulate these
increments from \(\theta_x=0\). We denote the resulting signed winding
by \(\nu_W\), consistently with the main text; in the adopted
orientation, \(\nu_W=C\). The Wilson curve and the Chern number thus
use the same determinant-link and plaquette data, with the curve
providing a complementary visualization of the bundle topology.

\subsection{Full-torus bundle validity}

The full two-dimensional validation is distinct from the
one-dimensional spectral-isolation diagnostic \(R_9^{(x)}\) shown
in Fig.~\ref{fig:bhz35}(d). On the complete sampled twist mesh
\(\mathcal T\), define
\begin{equation}
\begin{aligned}
\Delta_D^{\min}
&=
\min_{\boldsymbol\theta\in\mathcal T}
[E_D(\boldsymbol\theta)-E_{D-1}(\boldsymbol\theta)],\\
s_{\rm link}^{\min}
&=
\min_{\boldsymbol\theta\in\mathcal T,\,\mu}
s_{\min}[\mathbf M_\mu(\boldsymbol\theta)],
\end{aligned}
\label{eq:app_bundle_validity}
\end{equation}
where \(s_{\min}\) denotes the smallest singular value. An invariant
is accepted only after the complete mesh has been evaluated, the
low-energy candidate sets pass the completeness check, the sampled
external gap is finite and exceeds the prescribed threshold, and
every link satisfies \(s_{\min}\geq10^{-8}\) and
\(|\det\mathbf M_\mu|>10^{-14}\). The candidate-set check requires
the highest returned energy in each momentum sector to lie above
the top of the selected \(D\)-state manifold; at least \(D+1\)
candidate states are needed to evaluate the external gap. The gap
threshold is \(10^{-8}\) for the rank-three calculations on both
clusters and \(10^{-4}\) for the balanced \(3\times5\) rank-nine
calculations. The internal manifold width and gap-to-width ratio
characterize spectral isolation but are not additional acceptance
criteria.

Table~\ref{tab:app_bundle_validity} gives representative accepted
bundles, including the \(3\times5\) polarized check used to connect
the characterization in Sec.~\ref{subsec:bhz_polarized} to the common-cluster sector
comparison. Both balanced channels pass on complete \(5\times5\)
twist meshes at \(\lambda=0,0.1,\ldots,0.6\), with \(C_c=0\) and
\(C_{sc}=6\).

At the first sampled balanced-state failure, \(\lambda=0.7\), the
two channels fail differently. In the charge channel, the full mesh
has \(\Delta_9^{\min}\simeq6.79\times10^{-4}\), above the gap
threshold, but a neighboring-multiplet overlap is singular.
The mixed spin-charge calculation instead encounters a vanishing
external gap at \((\theta_x,\theta_y)=(2\pi/5,0)\) and stops there.
At \(\lambda=0.8,0.9,1.0\), both channels fail the gap criterion
at the first sampled twist. No valid rank-nine bundle invariant is
assigned when these criteria fail.
\section{Particle/hole entanglement spectra and quasihole counting}
\label{app:pes_counting}

\subsection{Entanglement construction and particle/hole cuts}

Following the standard particle-entanglement-spectrum
construction~\cite{SterdyniakEtAl2011}, we represent the rank-\(D\)
low-energy manifold at zero boundary twist by the basis-independent
equal-weight density matrix
\(\rho_D=D^{-1}\sum_{a=0}^{D-1}|\Psi_a\rangle\langle\Psi_a|\).
Tracing over the complementary particle subsystem \(B\) gives
\(\rho_A=\operatorname{Tr}_B\rho_D\), whose eigenvalues \(p_i\)
define the entanglement energies \(\xi_i=-\ln p_i\).

For the spectra in the main text, we retain \(N_A=3\) electrons for
the QWZ \(1/3\) state,
\((N_{A,\uparrow},N_{A,\downarrow})=(2,2)\) for the balanced BHZ
state, and \(N_A^h=3\) holes for the polarized \(2/3\) state.
The latter cut is performed in the hole representation relative
to the filled target band, treating the 20-electron state in
30 orbitals as a 10-hole state. This change of representation
does not assume particle--hole symmetry of the projected Hamiltonian.

\subsection{Quasihole counting and fixed-counting gap}

For a single fermionic Laughlin component, the torus
\((1,3)\)-admissibility rule described in
Sec.~\ref{subsec:qwz_entanglement} gives the quasihole reference count~\cite{WuBernevigRegnault2012,SterdyniakEtAl2011}
\begin{equation}
\mathcal N_{1/3}(N_\phi,N_A)
=
\frac{N_\phi}{N_\phi-2N_A}
\binom{N_\phi-2N_A}{N_A},
\label{eq:app_laughlin_counting}
\end{equation}
where \(N_A\) is the retained particle number in the electron or
hole representation. Thus \(\mathcal N_{1/3}(30,3)=2530\) for both
the QWZ three-electron cut and the polarized three-hole cut.
For the balanced BHZ cut, \(\mathcal N_{1/3}(15,2)=75\) for each
conserved component, giving the product-Laughlin reference count
\(75^2=5625\). At finite intercomponent coupling, this product count
is used as a reference for the low-lying entanglement branch and
does not assume an exactly factorized many-body wave function.

To evaluate the fixed-counting gap used in the main text, we pool
the entanglement levels from all subsystem-momentum sectors and
order them globally as \(\xi_1\leq\xi_2\leq\cdots\). For the
appropriate reference count \(N_{\rm qh}\) specified above, the gap is
\begin{equation}
\Delta_{\rm PES}
=
\xi_{N_{\rm qh}+1}-\xi_{N_{\rm qh}}.
\label{eq:app_pes_gap}
\end{equation}
The quasihole counting therefore fixes where the gap is evaluated
before the spectrum is inspected.
%
%
%
\section{Sector energetics and uniform pseudospin anisotropy}
\label{app:sector_energetics}

\subsection{Uniform anisotropy and sector-energy shifts}

For the conserved number sectors of Sec.~\ref{subsec:bhz_competition},
the uniform pseudospin anisotropy shifts all eigenvalues within a
sector equally. Using the polarization \(P\) and momentum-minimized
sector energy \(\mathcal E_P\) defined there, the resulting energy is
\begin{equation}
\begin{aligned}
\mathcal E_P(\lambda,g_z)
&=
\mathcal E_P(\lambda,0)
+\frac{N^2P^2}{4N_\phi}\,g_z\\
&=
\mathcal E_P(\lambda,0)
+\frac{5}{3}P^2g_z,
\end{aligned}
\label{eq:app_sector_shift}
\end{equation}
where the last equality applies to the \(N=10\), \(N_\phi=15\)
cluster used for the sector comparison. The factor \(1/N_\phi\)
makes the anisotropy contribution extensive at fixed filling
and polarization.

The projected interaction in Eq.~\eqref{eq:projected_hamiltonian}
omits the \(\mathbf q=0\) transfer, with the uniform total-density
Hartree contribution canceled by the neutralizing background.
In a multicomponent realization, however, the uniform total-density
and relative-component-density channels need not have the same
long-wavelength behavior. For layer- or orbital-resolved components,
the latter can retain a finite capacitive or pseudospin-anisotropy
energy after the common total-density contribution is neutralized~\cite{MoonEtAl1995,JungwirthMacDonald1996,JungwirthEtAl1998,JungwirthMacDonald2000,ShiEtAl2016}.
We use \(g_z\) to parametrize this uniform component-imbalance
energy independently of the finite-momentum intercomponent coupling
\(\lambda\). It is not assumed to be the literal \(\mathbf q\to0\)
continuation of the simplified Coulomb kernel in
Eq.~\eqref{eq:interaction}.

\subsection{Sector competition and ground-state envelope}

\begin{figure}[t]
\centering
\includegraphics[width=\columnwidth]{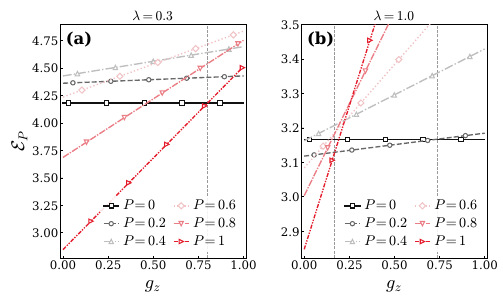}
\caption{
Sector ground-state energies \(\mathcal E_P(\lambda,g_z)\) on the
\(3\times5\) torus. The curves show the six conserved polarization
sectors \(P=0,0.2,\ldots,1\), obtained from their momentum-minimized
\(g_z=0\) energies through the uniform shift in
Eq.~\eqref{eq:app_sector_shift}. (a) At \(\lambda=0.3\), the lower
envelope changes directly from \(P=1\) to \(P=0\). (b) At
\(\lambda=1\), the \(P=0.2\) sector intervenes, giving
\(P_{\rm GS}=1\rightarrow0.2\rightarrow0\). Vertical dashed lines
mark changes of the lower envelope. The panels use independent
vertical ranges to resolve the crossings.
}
\label{fig:app_sector_energies}
\end{figure}

For \(N=10\), the allowed polarizations are \(P=0,0.2,\ldots,1\).
Opposite component imbalances are related by time reversal, so these
six values represent the independent sector energies. The numerical
input is the set of momentum-minimized energies
\(\mathcal E_P(\lambda,0)\) obtained by exact diagonalization.
Equation~\eqref{eq:app_sector_shift} then determines their \(g_z\)
dependence without further diagonalization. The global ground
sector is obtained from the lower envelope,
\begin{equation}
P_{\rm GS}(\lambda,g_z)
=
\operatorname*{arg\,min}_{P\in\{0,0.2,\ldots,1\}}
\mathcal E_P(\lambda,g_z).
\label{eq:app_ground_sector}
\end{equation}
Figure~\ref{fig:app_sector_energies} illustrates this construction
through the individual sector-energy curves underlying the
ground-sector selection in Fig.~\ref{fig:sector_competition}(b).

\bibliography{references}

\end{document}